\documentclass[pra,aps,preprint,amsmath,amssymb,showpacs]{revtex4}
\usepackage{graphicx}
\usepackage{epsfig}
\include{graphics}
\usepackage{epstopdf} 

\begin{document}

\title{Soliton-like behavior in fast two-pulse collisions 
in weakly perturbed linear physical systems}
\author{Avner Peleg$^{1}$, Quan M. Nguyen$^{2}$, and Toan T. Huynh$^{3}$}
\affiliation{$^{1}$Department of Exact Sciences, Afeka College of Engineering, 
Tel Aviv 69988, Israel}
\affiliation{$^{2}$Department of Mathematics, International University, 
Vietnam National University-HCMC, Ho Chi Minh City, Vietnam}
\affiliation{$^{3}$Department of Mathematics, University of Science, 
Vietnam National University-HCMC, Ho Chi Minh City, Vietnam}

\date{\today}
% updated : 01/18/2016
% This version is written using the dimensionless forms of the propagation models. 

\begin{abstract} 
We demonstrate that pulses of linear physical systems, 
weakly perturbed by nonlinear dissipation, 
exhibit soliton-like behavior in fast collisions. 
The behavior is demonstrated for linear waveguides 
with weak cubic loss and for systems described by 
linear diffusion-advection models with weak quadratic loss. 
We show that in both systems, the expressions for the collision-induced 
amplitude shifts due to the nonlinear loss have the same form as the expression 
for the amplitude shift in a fast collision between two optical solitons 
in the presence of weak cubic loss. 
Our analytic predictions are confirmed by 
numerical simulations with the corresponding coupled   
linear evolution models with weak nonlinear loss.   
These results open the way for studying dynamics of fast collisions  
between pulses of weakly perturbed linear physical systems 
in an arbitrary spatial dimension. 
\end{abstract}

%\pacs{42.65.Tg, 05.45.Yv, 47.35.Fg}
\maketitle
%\newpage
%\begin{multicols}{2}

\section{Introduction}
\label{Introduction}
Solitons, which are stable shape preserving traveling-wave solutions of nonlinear wave models, 
appear in a wide range of fields, including hydrodynamics \cite{Zakharov84}, 
condensed matter physics \cite{Malomed89}, optics \cite{Agrawal2001,Mollenauer2006}, 
and plasma physics \cite{Horton96}. The robustness of solitons in soliton collisions, 
i.e., the fact that the solitons do not change their shape in the collisions, 
is one of their most fundamental properties \cite{Zakharov84,Malomed89,Agrawal2001}. 
This property is often associated with the integrable nature of 
the corresponding nonlinear wave models \cite{Zakharov84}.

Another major property of solitons is manifested during {\it fast} inter-soliton 
collisions, i.e., during collisions for which the difference between the central 
frequencies (and group velocities) of the solitons is much larger than the soliton 
spectral width. This important property concerns the simple scaling relations   
satisfied by the soliton parameters, such as position, phase, amplitude, 
and frequency during fast collisions.  
It holds both in the absence of perturbations and in the presence 
of weak perturbations to the integrable nonlinear wave model. 
For example, the phase and position of fundamental solitons of the 
nonlinear Schr\"odinger (NLS) equation exhibit a shift during a two-soliton collision \cite{Zakharov84}. 
For fast collisions, the collision-induced phase and position shifts of soliton 1, 
for example, scale as $\eta_{2}/|\Delta\beta|$ and $-\eta_{2}/\Delta\beta^{2}$, 
where $\eta_{j}$ with $j=1,2$ are the soliton amplitudes, $\Delta\beta=\beta_{2}-\beta_{1}$, 
and $\beta_{j}$ with $j=1,2$ are the soliton frequencies \cite{Zakharov84,MM98,CP2005}. 
Furthermore, during fast collisions in the presence of weak cubic loss, 
solitons of the NLS equation experience 
amplitude and frequency shifts that scale as 
$-\epsilon_{3}\eta_{1}\eta_{2}/|\Delta\beta|$ and 
$-\epsilon_{3}\eta_{1}^{2}\eta_{2}/\Delta\beta^{2}$ for soliton 1, 
where $\epsilon_{3}$ is the cubic loss coefficient \cite{PNC2010}. 
Similar simple scaling relations hold for fast two-pulse collisions of NLS solitons in 
the presence of other weak perturbations, such as delayed Raman response 
\cite{Chi89,Malomed91,Kumar98,CP2005,P2004,NP2010}, 
and higher-order nonlinear loss \cite{PC2012}.

The simple scaling relations of collision-induced changes in soliton parameters 
are often associated with the shape preserving and stability properties of the 
solitons \cite{PNC2010,CP2005,PC2012}. 
Since the latter two properties are related with the integrability of the 
nonlinear wave model, one might also relate the simple scaling behavior in fast 
two-soliton collisions to the integrability of the model. In contrast, one expects 
very different behavior in collisions between pulses that are not shape preserving, 
since changes in pulse shape or instability might lead to the breakdown of the 
simple dynamics observed in fast two-soliton collisions. 
The latter expectation should certainly hold in linear physical systems 
that are weakly perturbed by nonlinear dissipation, since the pulses of the 
linear systems are in general not shape preserving \cite{Tkach97,Agrawal2001,Agrawal89a}. 
For this reason, it is often claimed that conclusions drawn from analysis of soliton 
collisions cannot be applied to collisions between pulses of weakly perturbed linear systems 
\cite{Agrawal2001,Mollenauer2006,Tkach97,Agrawal89b,PNC2010,PCG2003,PCG2004}.

In the current paper, we show that the point of view of fast collisions between 
pulses that are not shape preserving, described above, 
is erroneous. More specifically, we demonstrate that pulses of 
linear physical systems, weakly perturbed by nonlinear dissipation, exhibit 
simple {\it soliton-like} scaling behavior. The behavior is demonstrated for 
two major examples: (a) linear waveguide systems with weak cubic loss; 
(b) systems described by linear diffusion-advection models with weak 
quadratic loss. For both systems, we show that the expressions for the collision-induced 
amplitude shifts due to the nonlinear loss have the same form as the expression 
for the amplitude shift in a fast collision between two optical solitons 
in the presence of weak cubic loss. 
We validate our analytic predictions by numerical simulations 
with the corresponding perturbed coupled linear evolution models.   
Our results open the way for studying dynamics of fast collisions  
between pulses of weakly perturbed linear physical systems 
in an arbitrary spatial dimension, which is typically impossible for 
collisions between solitons in systems described by NLS models 
due to the instability of the solitons in dimension higher than one \cite{Zakharov84}.

The calculation of the collision-induced amplitude shift in the current paper 
is based on a generalization of the the perturbation technique, developed in Refs. 
\cite{PCG2003,PCG2004} for calculating the effects of weak perturbations 
on fast collisions between NLS solitons. This perturbation technique was first 
used to calculate the effects of weak conservative perturbations, such as 
third-order dispersion \cite{PCG2003,PCG2004} and quintic nonlinearity \cite{SP2004} 
on fast two-soliton collisions. Later on it was shown that the perturbation technique 
can also be used for calculating the effects of weak dissipative perturbations on 
fast soliton collisions \cite{CP2005,PNC2010,PC2012,P2004}. 
In the current paper we further generalize and extend the perturbation technique 
to allow treatment of fast two-pulse collisions in linear physical systems, weakly perturbed 
by nonlinear dissipation.  The main assumption of the generalized perturbation 
technique is that the smallest relevant length scale (or time scale) in the problem is the 
collision length (or collision time interval), which is the distance (or time interval) 
along which the two colliding pulses overlap. This assumption along with the 
assumptions of a fast collision and weak dissipation allow us to obtain simple 
scaling relations for the collision-induced amplitude shifts, which are similar in 
form to the simple scaling relations obtained for fast collisions between two 
optical solitons in the presence of weak cubic loss.

The rest of the paper is organized as follows. 
In section \ref{waveguides}, we obtain the analytic prediction for the collision-induced amplitude 
shift in a fast collision between two optical pulses in a linear waveguide with 
weak linear and cubic loss. We then compare the analytic prediction with results 
of numerical simulations of the collision with the perturbed coupled linear propagation 
model. In section \ref{diffusion}, we obtain the analytic prediction for the collision-induced amplitude 
shift in a fast collision between two concentration pulses in systems described by 
perturbed coupled linear diffusion-advection models with weak linear and quadratic loss. 
In addition, we present a comparison of the analytic prediction with the results of numerical 
simulations with the perturbed coupled linear diffusion-advection model.  
In section \ref{conclusions}, we present our conclusions. 
In Appendix \ref{appendA}, we derive the relations between the collision-induced 
amplitude shift and the collision-induced change in pulse shape. 
Appendix \ref{appendB} is devoted to a description of the procedures used 
for calculating the values of the collision-induced amplitude shift from 
the analytic predictions and from results of numerical simulations.

\section{Fast collisions in linear waveguides}
\label{waveguides} 
\subsection{Propagation model}
\label{waveguides_model} 
We consider fast collisions between two optical pulses in linear waveguides with 
weak linear and cubic loss. The dynamics of the collision is described by   
the following system of perturbed coupled linear propagation equations \cite{Agrawal2001,Agrawal2007a,PNC2010}: 
\begin{eqnarray}&&
\!\!\!\!\!\!\!
i\partial_z\psi_{1}\!-\!\mbox{sgn}(\tilde\beta_{2})\partial_{t}^{2}\psi_{1}\!=\!
-i\epsilon_{1}\psi_{1}\!
-\!i\epsilon_{3}|\psi_{1}|^2\psi_{1}\!
-\!2i\epsilon_{3}|\psi_{2}|^2\psi_{1},
\nonumber \\&&
\!\!\!\!\!\!\!
i\partial_z\psi_{2}+id_{1}\partial_{t}\psi _{2}
-\mbox{sgn}(\tilde\beta_{2})\partial_{t}^2\psi_{2}=
-i\epsilon_{1}\psi_{2}-i\epsilon_{3}|\psi_{2}|^2\psi_{2}
\nonumber \\&&
-2i\epsilon_{3}|\psi_{1}|^2\psi_{2},  
\!\!\!\!\!\!\!\!
\label{coll1}
\end{eqnarray}           
where $\psi_{1}$ and $\psi_{2}$ are the envelopes of the electric fields of the pulses, 
$z$ is propagation distance, and $t$ is time \cite{Dimensions1}. 
In Eq. (\ref{coll1}), $d_{1}$ is the group velocity coefficient, 
$\tilde\beta_{2}$ is the second-order dispersion coefficient, and 
$\epsilon_{1}$ and $\epsilon_{3}$ are the linear and cubic loss coefficients, 
which satisfy $0<\epsilon_{1} \ll 1$ and $0<\epsilon_{3} \ll 1$.   
The terms $-\mbox{sgn}(\tilde\beta_{2})\partial_{t}^{2}\psi_{j}$ 
on the left hand side of Eq. (\ref{coll1}) are due to second-order dispersion, 
while $id_{1}\partial_{t}\psi _{2}$ is associated with the group velocity difference. 
The first terms on the right hand side of Eq. (\ref{coll1}) describe linear loss effects, 
while the second and third terms describe intra-pulse and inter-pulse effects 
due to cubic loss.

\subsection{Calculation of the amplitude shift in a fast two-pulse collision}        
\label{waveguides_delta_A} 
We consider a fast collision between two pulses with generic shapes 
and with tails that exhibit exponential or faster than exponential decay. 
We assume that the pulses can be characterized by initial amplitudes $A_{j}(0)$, 
initial widths $W_{j0}$, initial positions $y_{j0}$, and initial phases $\alpha_{j0}$. 
Thus, for a collision between two Gaussian pulses, 
for example, the initial envelopes of the electric fields are: 
\begin{eqnarray}&&
\!\!\!\!\!\!\!
\psi_{j}(t,0)=A_{j}(0)
\exp[-(t-y_{j0})^{2}/(2W_{j0}^{2})+i\alpha_{j0}],  
\!\!\!\!\!\!\!\!
\label{IC1}
\end{eqnarray}       
where $j=1,2$. As another example, for a collision between two square pulses, 
the initial envelopes of the electric fields are:  
\begin{eqnarray} &&
\!\!\!\!\!\!\!\!\!
\psi_{j}(t,0)=
\left\{\begin{array}{l l}
A_{j}(0)\exp(i\alpha_{j0}) & \;\; \mbox{for} \;\;  |t-y_{j0}| \le W_{j0}/2,\\
0 & \;\; \mbox{for} \;\; |t-y_{j0}| > W_{j0}/2,\\
\end{array} \right. 
%\nonumber \\&&
\label{IC2}
\end{eqnarray} 
where $j=1,2$. We assume a complete collision, 
such that the two pulses are well separated at $z=0$ 
and at the final distance $z=z_{f}$.

Let us discuss the implications of the assumption of a fast collision. 
For this purpose, we define the collision length $\Delta z_{c}$, 
which is the distance along which the envelopes of the colliding pulses overlap, 
by $\Delta z_{c}=W_{0}/|d_{1}|$, where for simplicity we assume 
$W_{10}=W_{20}=W_{0}$. The assumption of a fast collision means that 
$\Delta z_{c}$ is the shortest length scale in the problem. 
In particular, $\Delta z_{c}\ll z_{D}$, where $z_{D}=W_{0}^{2}/2$ 
is the dispersion length. Using the definitions of $\Delta z_{c}$ and $z_{D}$, 
we obtain $W_{0}|d_{1}|/2 \gg 1$, as the condition for a fast collision.

Our perturbative calculation of the amplitude shift in a fast collision is 
a generalization of the perturbative technique, developed in Refs. 
\cite{PCG2003,PCG2004} for calculating the effects of weak perturbations 
on fast two-soliton collisions. In an analogy with the fast two-soliton collision 
case,  we look for a solution of Eq. (\ref{coll1}) in the form 
\begin{eqnarray}&&
\!\!\!\!\!\!\!
\psi_{j}(t,z)=\psi_{j0}(t,z)+\phi_{j}(t,z), 
\label{coll2}
\end{eqnarray}       
where $j=1,2$, $\psi_{j0}$ are the solutions of Eq. (\ref{coll1}) without 
the inter-pulse interaction terms, and $\phi_{j}$ describe corrections to 
$\psi_{j0}$ due to inter-pulse interaction. By definition, 
$\psi_{10}$ and $\psi_{20}$ satisfy 
\begin{eqnarray}&&
\!\!\!\!\!\!\!
i\partial_z\psi_{10}\!-\!\mbox{sgn}(\tilde\beta_{2})\partial_{t}^{2}\psi_{10}\!=\!
-i\epsilon_{1}\psi_{10}\!
-\!i\epsilon_{3}|\psi_{10}|^2\psi_{10},
\!\!\!\!\!\!\!\!
\label{coll2_add1}
\end{eqnarray}         
and
\begin{eqnarray}&&
\!\!\!\!\!\!\!
i\partial_z\psi_{20}+id_{1}\partial_{t}\psi _{20}
-\mbox{sgn}(\tilde\beta_{2})\partial_{t}^2\psi_{20}=
\nonumber \\&&
-i\epsilon_{1}\psi_{20}
-i\epsilon_{3}|\psi_{20}|^2\psi_{20}.
%\!\!\!\!\!\!\!\!
\label{coll2_add2}
\end{eqnarray}         
We now substitute relation (\ref{coll2}) into Eq. (\ref{coll1}) 
and use Eqs. (\ref{coll2_add1}) and  (\ref{coll2_add2}) 
to obtain equations for the $\phi_{j}$. We concentrate 
on the calculation of $\phi_{1}$, since the calculation of $\phi_{2}$ is similar. 
Taking into account only leading-order effects of the collision, 
we can neglect terms containing $\phi_{j}$ on the right hand side 
of the resulting equation. We therefore obtain: 
\begin{equation}
i\partial_z\phi_{1}-\mbox{sgn}(\tilde\beta_{2})\partial_{t}^{2}\phi_{1}=
-2i\epsilon_{3}|\psi_{20}|^2\psi_{10}.
\label{coll3} 
\end{equation}      
Continuing the analogy with the fast two-soliton collision, 
we substitute $\psi_{j0}(t,z)=\Psi_{j0}(t,z)\exp[i\chi_{j0}(t,z)]$ and 
$\phi_{1}(t,z)=\Phi_{1}(t,z)\exp[i\chi_{10}(t,z)]$, where $\Psi_{j0}$ and 
$\chi_{j0}$ are real-valued, into Eq. (\ref{coll3}).
This substitution yields the following equation for $\Phi_{1}$: 
\begin{eqnarray} &&
i\partial_{z}\Phi_{1} - \left(\partial_{z}\chi_{10}\right)\Phi_{1}
-\mbox{sgn}(\tilde\beta_{2})\left[ 
\partial_{t}^{2}\Phi_{1}
+2i\left(\partial_{t}\chi_{10}\right)\partial_{t}\Phi_{1}
\right. 
\nonumber \\&&
\left. 
+i\left(\partial_{t}^{2}\chi_{10}\right)\Phi_{1}
-\left(\partial_{t}\chi_{10}\right)^2\Phi_{1}
\right] =
-2\epsilon_{3}\Psi_{20}^{2}\Psi_{10}.
\label{coll3_add1}
\end{eqnarray} 
The term on the right hand side  of Eq. (\ref{coll3_add1}) is of order $\epsilon_{3}$. 
In addition, since the collision length $\Delta z_{c}$ is of order 
$1/|d_{1}|$, the term $i\partial_{z}\Phi_{1}$ is of order 
$|d_{1}| \times O(\Phi_{1})$. Equating the orders of $i\partial_{z}\Phi_{1}$ 
and $-2\epsilon_{3}\Psi_{20}^{2}\Psi_{10}$, we find that $\Phi_{1}$ 
is of order $\epsilon_{3}/|d_{1}|$. In addition, we observe that 
all other terms on the left hand side of Eq. (\ref{coll3_add1}) are 
of order $\epsilon_{3}/|d_{1}|$ or higher, and can therefore be neglected. 
As a result, in the leading order of the perturbative calculation, 
the equation for the collision-induced change in the envelope of pulse 1 is:
\begin{eqnarray} &&
\partial_{z}\Phi_{1}=
-2\epsilon_{3}\Psi_{20}^{2}\Psi_{10}.
\label{coll4}
\end{eqnarray} 
Equation (\ref{coll4}) is similar to the equation obtained 
for a fast collision between two optical solitons in a nonlinear optical waveguide 
with weak cubic loss (see Eq. (9) in Ref. \cite{PNC2010}).

The collision-induced amplitude shift of pulse 1 is calculated from  
the collision-induced change in the envelope of pulse 1. 
We denote by $z_{c}$ the collision distance, 
which is the distance at which the maxima of $|\psi_{j}(t,z)|$ coincide.  
In a fast collision, the collision takes place in a small 
interval $[z_{c}-\Delta z_{c},z_{c}+\Delta z_{c}]$ about $z_{c}$.  
Therefore, the net collision-induced change in the envelope of pulse 1 
$\Delta\Phi_{1}(t,z_{c})$ can be evaluated by: 
$\Delta\Phi_{1}(t,z_{c})=\Phi_{1}(t,z_{c}+
\Delta z_{c})-\Phi_{1}(t,z_{c}-\Delta z_{c})$. 
To calculate $\Delta\Phi_{1}(t,z_{c})$,    
we use the approximation: $\Psi_{j0}(t,z)=A_{j}(z)\tilde\Psi_{j0}(t,z)$, 
where $\tilde\Psi_{j0}(t,z)\exp[i\chi_{j0}(t,z)]$ is the solution 
of the propagation equation without linear and cubic loss
and with $A_{j}(0)=1$. Substituting the approximate 
expressions for $\Psi_{j0}$ into Eq. (\ref{coll4}) 
and integrating with respect to $z$ over the interval 
$[z_{c}-\Delta z_{c},z_{c}+\Delta z_{c}]$, we obtain: 
\begin{eqnarray}&&
\!\!\!\!\!\!
\Delta\Phi_{1}(t,z_{c})\!=\!-2\epsilon_{3}
\!\!\int_{z_{c}-\Delta z_{c}}^{z_{c}+\Delta z_{c}} 
\!\!\!\!\!\!\!\!\!\!\! dz' A_{1}(z') A_{2}^{2}(z')
\tilde\Psi_{10}(t,z')\tilde\Psi_{20}^{2}(t,z').
\nonumber \\&&
\label{coll5}
\end{eqnarray}  
The only function on the right hand side of Eq. (\ref{coll5}) 
that contains fast variations in $z$, which are of order 1, is $\tilde\Psi_{20}$. 
As a result, we can approximate $A_{1}(z)$, $A_{2}(z)$, and $\tilde\Psi_{10}(t,z)$ 
by $A_{1}(z_{c}^{-})$, $A_{2}(z_{c}^{-})$, and $\tilde\Psi_{10}(t,z_{c})$, 
where $A_{j}(z_{c}^{-})$ denotes the limit from the left of $A_{j}$ at $z_{c}$. 
Furthermore, in calculating the integral we can take into account only the fast dependence 
of $\tilde\Psi_{20}$ on $z$, i.e., the $z$ dependence that is contained in factors of the form 
$y=t-y_{20}-d_{1}z$. Denoting this approximation of 
$\tilde\Psi_{20}(t,z)$ by $\bar\Psi_{20}(y,z_{c})$, we obtain: 
\begin{eqnarray}&&
\!\!\!\!\!\!\!\!
\Delta\Phi_{1}(t,z_{c})\!=\!
-2\epsilon_{3}A_{1}(z_{c}^{-}) A_{2}^{2}(z_{c}^{-})
\tilde\Psi_{10}(t,z_{c})\times
\nonumber \\&&
\!\!\int_{z_{c}-\Delta z_{c}}^{z_{c}+\Delta z_{c}}  
\!\!\!\!\!\!\!\! dz' 
\bar\Psi_{20}^{2}(t-y_{20}-d_{1}z',z_{c}).
\label{coll5_add1}
\end{eqnarray}    
Since the integrand on the right hand side of Eq. (\ref{coll5_add1}) 
is sharply peaked at a small interval about $z_{c}$,  
we can extend the limits of the integral to $-\infty$ and $\infty$. 
We also change the integration variable from 
$z'$ to $y=t-y_{20}-d_{1}z'$ and obtain: 
\begin{eqnarray} &&
% one-line version 1
\!\!\!\!\!\!\!
\Delta\Phi_{1}(t,z_{c})\!=\!-\frac{2\epsilon_{3}
A_{1}(z_{c}^{-}) A_{2}^{2}(z_{c}^{-})}{|d_{1}|}\tilde\Psi_{10}(t,z_{c})
\!\!\!\int_{-\infty}^{\infty} \!\!\!\!\!\!\!\!\! dy \bar\Psi_{20}^{2}(y,z_{c}).
\nonumber \\&&
\label{coll6}
\end{eqnarray}
The total collision-induced amplitude shift of pulse 1 $\Delta A_{1}^{(c)}$ 
is related to the net collision-induced change in the envelope of the pulse  
$\Delta\Phi_{1}(t,z_{c})$ by:
\begin{eqnarray}&&
\!\!\!\!\!\!\!\!\!\!\!\!\!\!
\Delta A_{1}^{(c)}=
\left[\int_{-\infty}^{\infty} \!\!\!\!\! dt \tilde\Psi_{10}^{2}(t,z_{c})\right]^{-1}
\!\!\int_{-\infty}^{\infty} \!\!\!\!\! dt \tilde\Psi_{10}(t,z_{c})\Delta\Phi_{1}(t,z_{c}) 
\label{coll6_add1}
\end{eqnarray}       
(see Appendix \ref{appendA}). Substituting Eq. (\ref{coll6}) into  Eq. (\ref{coll6_add1}),  
we find that the total collision-induced amplitude shift of pulse 1 is: 
\begin{eqnarray} &&
\!\!\!\!
\Delta A_{1}^{(c)}=-\frac{2\epsilon_{3}
A_{1}(z_{c}^{-}) A_{2}^{2}(z_{c}^{-})}{|d_{1}|}
\int_{-\infty}^{\infty} dy \bar\Psi_{20}^{2}(y,z_{c}).
\label{coll7}
\end{eqnarray}    
Equation (\ref{coll7}) is expected to hold for generic pulse shapes $\Psi_{j0}(t,z)$
with tails that exhibit exponential or faster than exponential decay.
Indeed, in this case the approximations leading from 
Eq. (\ref{coll5}) to Eq. (\ref{coll6}) are expected to be valid. 
Employing Eq. (\ref{coll7}) for a fast collision between two Gaussian pulses 
with initial widths $W_{j0}$, we find that the collision-induced amplitude shift 
in this case is given by: 
\begin{eqnarray} &&
\Delta A_{1}^{(c)}=
-2 \pi^{1/2} \epsilon_{3} W_{20}
A_{1}(z_{c}^{-}) A_{2}^{2}(z_{c}^{-})/|d_{1}|.
\label{coll8}
\end{eqnarray}  
In a similar manner, using Eq. (\ref{coll7}) we find that the collision-induced 
amplitude shift in a fast collision between two square pulses with initial widths 
$W_{j0}$ is given by: 
\begin{eqnarray} &&
\Delta A_{1}^{(c)}=
-2 \epsilon_{3} W_{20}
A_{1}(z_{c}^{-}) A_{2}^{2}(z_{c}^{-})/|d_{1}|.
\label{coll8_add1}
\end{eqnarray}                          
Expressions (\ref{coll7})-(\ref{coll8_add1}) are very similar 
to the expression obtained in Ref. \cite{PNC2010} for the amplitude shift 
in a fast collision between two optical solitons in the presence of weak cubic loss: $\Delta\eta_{1}^{(c)}=-4\epsilon_{3}\eta_{1}(z_{c}^{-})\eta_{2}(z_{c}^{-})/|\Delta\beta|$
(see Eq. (11) in Ref. \cite{PNC2010}). Indeed, since the soliton width is $W_{j}=1/\eta_{j}$, 
we can express the collision-induced amplitude shift of the soliton as:
\begin{eqnarray} &&
\Delta\eta_{1}^{(c)}=
-4\epsilon_{3}W_{2}(z_{c}^{-})\eta_{1}(z_{c}^{-})
\eta_{2}^{2}(z_{c}^{-})/|\Delta\beta| .
\label{coll8_add2}
\end{eqnarray}      
Based on the similarity between Eqs. (\ref{coll7})-(\ref{coll8_add1}) 
and Eq. (\ref{coll8_add2}) we conclude that pulses of the linear propagation 
equation exhibit soliton-like behavior in fast collisions, and that this    
behavior is not sensitive to the pulse shape details.

\subsection{Numerical simulations}
\label{waveguides_simu} 
To validate the predictions for soliton-like behavior in two-pulse collisions, 
we carry out numerical simulations with Eq. (\ref{coll1}). The equation is 
numerically integrated by employing the split-step method 
with periodic boundary conditions \cite{Agrawal2001}. 
For concreteness, we present the results of  
simulations with parameter values $\epsilon_{1}=0.01$, $\epsilon_{3}=0.01$, 
and $\mbox{sgn}(\beta_{2})=1$. The values of $d_{1}$ are varied 
in the intervals $-60 \le d_{1} \le -4$ and $4 \le d_{1} \le 60$.      
We illustrate the behavior of the collision-induced amplitude shift for two 
different initial conditions, one corresponding to a collision between two 
Gaussian pulses, and the other corresponding to a collision between 
two square pulses. The initial condition for the first set of simulations 
consists of two Gaussian pulses of the form (\ref{IC1}) with parameter values 
$A_{j}(0)=1$, $W_{j0}=4$, $y_{10}=0$, $y_{20}=\pm 20$, 
and $\alpha_{j0}=0$. The initial condition for the second set of simulations 
consists of two square pulses of the form (\ref{IC2}) with parameter values 
$A_{j}(0)=1$, $W_{j0}=4$, $y_{10}=0$, $y_{20}=\pm 5$, 
and $\alpha_{j0}=0$. The procedures used for obtaining the values of 
$\Delta A_{1}^{(c)}$ from Eqs. (\ref{coll8}) and (\ref{coll8_add1}) and 
for calculating $\Delta A_{1}^{(c)}$ from the results of the numerical simulations 
are described in Appendix \ref{appendB}.

Figure \ref{fig1} shows the dependence of the collision-induced 
amplitude shift $\Delta A_{1}^{(c)}$ on $d_{1}$ for fast collisions 
between two Gaussian pulses. Both the result obtained by simulations
with Eq. (\ref{coll1}) and the analytic prediction of Eq. (\ref{coll8}) are shown. 
It is seen that the agreement between the simulations 
and the analytic prediction is very good. 
More specifically, the relative error in the approximation, which is defined by 
$|\Delta A_{1}^{(c)(num)}-\Delta A_{1}^{(c)(th)}|\times 100/
|\Delta A_{1}^{(c)(num)}|$, is less than 10$\%$ for $|d_{1}|>10$ 
and less than 2$\%$ for $|d_{1}|>20$. Even at $|d_{1}| \simeq 4$, 
the relative error is less than 25$\%$.

\begin{figure}[htbp]
\centerline{\includegraphics[width=0.95\columnwidth]{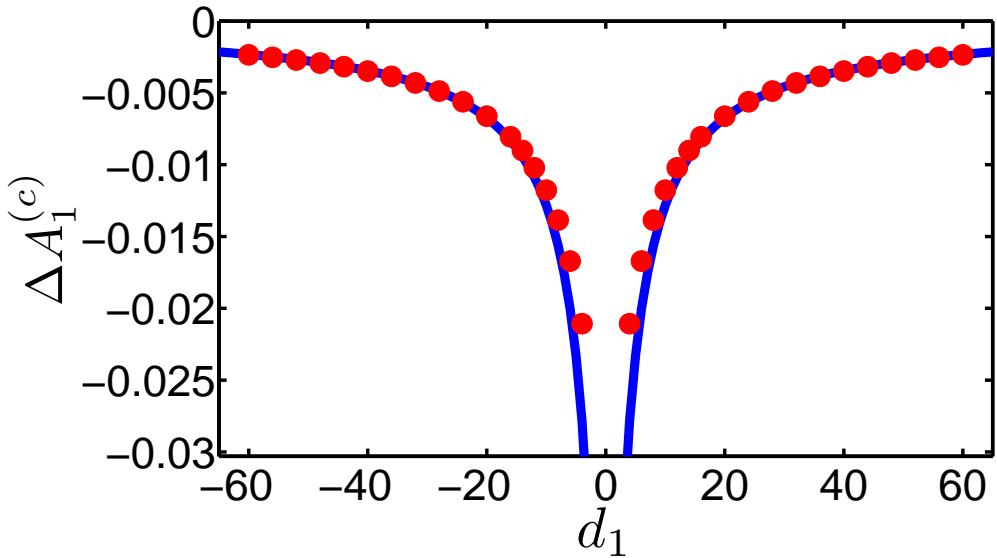}}
\caption{(Color online) The collision-induced amplitude shift of pulse 1 
$\Delta A_{1}^{(c)}$ vs group velocity parameter $d_{1}$ 
in a fast collision between two Gaussian pulses in a linear waveguide 
with weak linear and cubic loss. 
The red circles correspond to the result obtained by  
simulations with Eq. (\ref{coll1}). The solid blue line corresponds 
to the prediction of Eq. (\ref{coll8}).}
\label{fig1}
\end{figure}

\begin{figure}[htbp]
\centerline{\includegraphics[width=0.95\columnwidth]{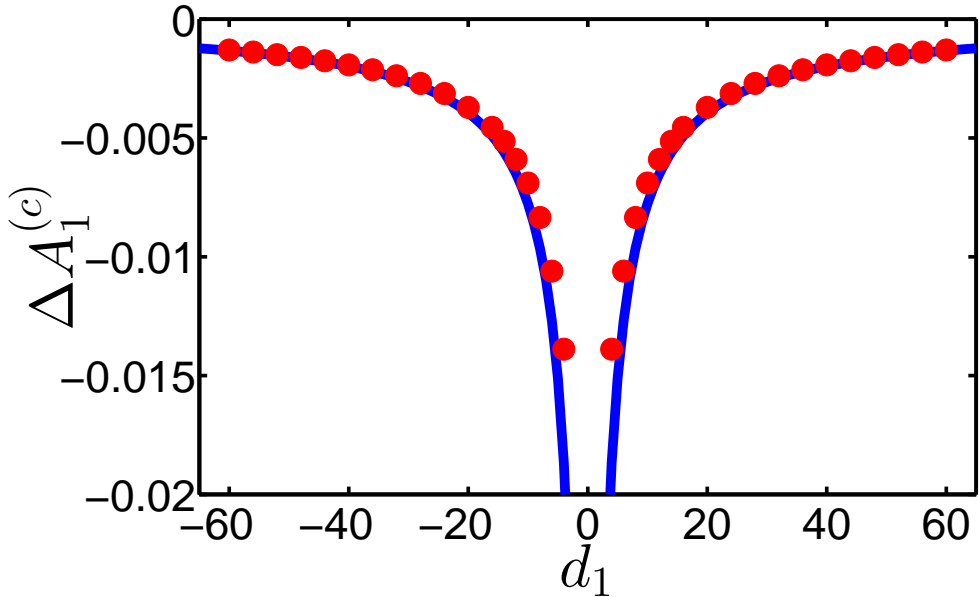}}
\caption{(Color online) The collision-induced amplitude shift of pulse 1 
$\Delta A_{1}^{(c)}$ vs group velocity parameter $d_{1}$ 
in a fast collision between two square pulses in a linear waveguide 
with weak linear and cubic loss. 
The red circles correspond to the result obtained by  
simulations with Eq. (\ref{coll1}). The solid blue line corresponds 
to the prediction of Eq. (\ref{coll8_add1}).}
\label{fig2}
\end{figure}

Figure \ref{fig2} shows the dependence of $\Delta A_{1}^{(c)}$ on $d_{1}$ 
for fast collisions between two square pulses, as obtained by numerical  simulations
with Eq. (\ref{coll1}). The analytic prediction of Eq.  (\ref{coll8_add1}) 
is also shown. The agreement between the numerical simulations and 
the analytic prediction is very good. In particular, the relative error 
in the approximation is less than 11$\%$ for $|d_{1}|>10$ 
and less than 6$\%$ for $|d_{1}|>20$. At $|d_{1}|\simeq 4$,  
the relative error is 26$\%$. Similar results to the ones presented 
in Figs. \ref{fig1} and \ref{fig2} are obtained for other choices 
of the physical parameter values and for other pulse shapes. 
We therefore conclude that pulses of the linear 
propagation equation indeed exhibit soliton-like behavior in fast 
collisions in the presence of weak cubic loss.

\section{Fast collisions in systems described by coupled linear 
diffusion-advection models}
\label{diffusion}
\subsection{Evolution model}
\label{diffusion_model}
We now turn to describe the dynamics of fast collisions between pulses 
of two substances, denoted by 1 and 2, that evolve in the presence of 
linear diffusion and weak linear and quadratic loss. In addition, we assume 
that material 2 is advected with velocity $v_{d}$ relative to material 1.  
The dynamics of the two-pulse collision is described by   
the following system of perturbed coupled linear diffusion-advection equations: 
\begin{eqnarray} &&
\!\!\!\!\!\!\!\!\!\!\!\!\!\!
\partial _{t}u_{1}=\partial _{x}^{2}u_{1}-\varepsilon_{1}u_{1}
-\varepsilon_{2}u_{1}^{2}-2\varepsilon_{2}u_{1}u_{2},
\nonumber \\&&
\!\!\!\!\!\!\!\!\!\!\!\!\!\!
\partial _{t}u_{2}=\partial _{x}^{2}u_{2}-v_{d}\partial _{x}u_{2}
-\varepsilon_{1}u_{2}-\varepsilon_{2}u_{2}^{2}-2\varepsilon_{2}u_{1}u_{2},
\label{rda1}
\end{eqnarray}         
where $u_{1}$ and $u_{2}$ are the concentrations of substance 1 and 2,  
$t$ is time, $x$ is a spatial coordinate, and $\varepsilon_{1}$ and $\varepsilon_{2}$ 
are the linear and quadratic loss coefficients, 
which satisfy $0<\varepsilon_{1} \ll 1$ and $0<\varepsilon_{2} \ll 1$ \cite{Dimensions2}.     
The term $-v_{d}\partial _{x}u_{2}$ in Eq. (\ref{rda1}) describes advection, 
while the terms $-\varepsilon_{1}u_{j}$ correspond to linear loss. 
The terms $-\varepsilon_{2}u_{j}^{2}$ and $-2\varepsilon_{2}u_{j}u_{k}$ 
describe intra-substance and inter-substance effects due to quadratic loss, respectively.

\subsection{Calculation of the amplitude shift in a fast two-pulse collision}  
\label{diffusion_delta_A}
We consider a fast collision between two pulses 
of substances 1 and 2 with generic shapes 
and with tails that exhibit exponential or faster than exponential decay. 
We assume that the pulses can be characterized by
initial amplitudes $A_{j}(0)$, initial widths $W_{j0}$, and initial positions $x_{j0}$. 
Therefore, for a collision between two Gaussian pulses, 
for example, the initial concentrations are: 
\begin{eqnarray}&&
\!\!\!\!\!\!\!
u_{j}(x,0)=A_{j}(0)\exp[-(x-x_{j0})^{2}/(2W_{j0}^{2})],  
\!\!\!\!\!\!\!\!
\label{rda_IC1}
\end{eqnarray}     
where $j=1,2$. Additionally, for a collision between two square pulses,  
the initial concentrations are: 
\begin{eqnarray} &&
\!\!\!\!\!\!\!\!\!
u_{j}(x,0)=
\left\{\begin{array}{l l}
A_{j}(0) & \;\; \mbox{for} \;\;  |x-x_{j0}| \le W_{j0}/2,\\
0 & \;\; \mbox{for} \;\; |x-x_{j0}| > W_{j0}/2,\\
\end{array} \right. 
%\nonumber \\&&
\label{rda_IC2}
\end{eqnarray} 
where $j=1,2$. We assume a complete collision, that is, a collision in which  
the pulses are well separated at $t=0$ and at the final time $t=t_{f}$.
The assumption of a fast collision means that the time interval 
$\Delta t_{c}=W_{0}/|v_{d}|$, along which the two pulses overlap, 
is much shorter than the diffusion time $t_{D}=W_{0}^{2}$.   
Requiring $\Delta t_{c} \ll t_{D}$, we obtain $W_{0}|v_{d}| \gg 1$, 
as the condition for a fast collision.

The perturbative calculation of the collision-induced amplitude shift is 
similar to the one carried out in section \ref{waveguides_delta_A} 
for fast collisions in linear waveguides with weak cubic loss.    
Thus, we look for a solution of Eq. (\ref{rda1}) in the form 
\begin{eqnarray}&&
\!\!\!\!\!\!\!
u_{j}(x,t)=u_{j0}(x,t)+\phi_{j}(x,t), 
\label{rda2}
\end{eqnarray}      
where $j=1,2$, $u_{j0}$ are solutions of Eq. (\ref{rda1}) 
without inter-pulse interaction, and $\phi_{j}$ describe collision-induced effects. 
By definition, $u_{10}$ and $u_{20}$ satisfy the equations  
\begin{eqnarray}&&
\!\!\!\!\!\!\!
\partial _{t}u_{10}=\partial _{x}^{2}u_{10}-\varepsilon_{1}u_{10}
-\varepsilon_{2}u_{10}^{2},
\!\!\!\!\!\!\!\!
\label{rda2_add1}
\end{eqnarray}         
and 
\begin{eqnarray}&&
\!\!\!\!\!\!\!
\partial _{t}u_{20}=\partial _{x}^{2}u_{20}-v_{d}\partial _{x}u_{20}
-\varepsilon_{1}u_{20}-\varepsilon_{2}u_{20}^{2}.
\!\!\!\!\!\!\!\!
\label{rda2_add2}
\end{eqnarray}         
We substitute relation (\ref{rda2}) into Eq. (\ref{rda1}) 
and use Eqs. (\ref{rda2_add1}) and  (\ref{rda2_add2}) 
to obtain equations for $\phi_{1}$ and $\phi_{2}$.
We concentrate on the calculation of $\phi_{1}$, as the calculation of $\phi_{2}$ is similar.     
Taking into account only leading-order effects of the collision, 
we can neglect terms of the form $-\varepsilon_{1}\phi_{1}$, 
$-2\varepsilon_{2}u_{10}\phi_{1}$, $-2\varepsilon_{2}u_{20}\phi_{1}$, 
$-2\varepsilon_{2}u_{10}\phi_{2}$, etc. We therefore obtain: 
\begin{equation}
\partial_{t}\phi_{1}=\partial _{x}^{2}\phi_{1}-2\varepsilon_{2}u_{10}u_{20}. 
\label{rda3} 
\end{equation}         
The term $-2\varepsilon_{2}u_{10}u_{20}$ on the right hand side 
of Eq. (\ref{rda3}) is of order $\varepsilon_{2}$. 
Equating the orders of $\partial_{t}\phi_{1}$ and $-2\varepsilon_{2}u_{10}u_{20}$ 
and taking into account that $\Delta t_{c}$ is of order $1/|v_{d}|$, 
we find that $\phi_{1}$ is of order $\varepsilon_{2}/|v_{d}|$. 
In addition, the term  $\partial _{x}^{2}\phi_{1}$, which is of order 
$\varepsilon_{2}/|v_{d}|$, can be neglected. 
Therefore, in the leading order, the equation 
for the collision-induced change of pulse 1 is:    
\begin{equation}
\partial _{t}\phi_{1}=-2\varepsilon_{2}u_{10}u_{20}.
\label{rda4} 
\end{equation}           
Equation  (\ref{rda4}) is similar to Eq. (\ref{coll4}) 
and also to the equation obtained in Ref. \cite{PNC2010} 
for a fast collision between two optical solitons 
in a nonlinear optical waveguide with weak cubic loss.

We calculate the collision-induced amplitude shift of pulse 1 from  
the collision-induced change in the concentration of pulse 1. 
For this purpose, we denote by $t_{c}$ the collision time, 
i.e., the time at which the maxima of $u_{j}(x,t)$ coincide.  
In a fast collision, the collision takes place in a small 
time interval $[t_{c}-\Delta t_{c},t_{c}+\Delta t_{c}]$ about $t_{c}$.  
Therefore, the net collision-induced change in the concentration of pulse 1 
$\Delta\phi_{1}(x,t_{c})$ can be evaluated by: 
$\Delta\phi_{1}(x,t_{c})=\phi_{1}(x,t_{c}+\Delta t_{c})-
\phi_{1}(x,t_{c}-\Delta t_{c})$. To calculate $\Delta\phi_{1}(x,t_{c})$,    
we use the approximation: $u_{j0}(x,t)=A_{j}(t) \tilde u_{j0}(x,t)$, 
where $\tilde u_{j0}(x,t)$ is the solution 
of the diffusion equation without linear and quadratic loss and with $A_{j}(0)=1$.
Substituting the approximate expressions for $u_{j0}$ into Eq. (\ref{rda4}) 
and integrating with respect to time over the interval 
$[t_{c}-\Delta t_{c},t_{c}+\Delta t_{c}]$, we obtain: 
\begin{eqnarray} &&
\!\!\!\!\!\!\!\!\!\!\!
\Delta\phi_{1}(x,t_{c})\!=\!
-2\varepsilon_{2}\!\!\int_{t_{c}-\Delta t_{c}}^{t_{c}+\Delta t_{c}} 
\!\!\!\!\!\!\!\!\!\!\!\!\!\!\!\! dt' A_{1}(t') A_{2}(t')
\tilde u_{10}(x,t') \tilde u_{20}(x,t'). 
\nonumber \\&&
\label{rda5}
\end{eqnarray}  
The only function on the right hand side of Eq. (\ref{rda5}) 
that contains fast variations in $t$, which are of order 1, is $\tilde u_{20}$. 
Therefore, we can approximate $A_{1}(t)$, $A_{2}(t)$, 
and $\tilde u_{10}(x,t)$ by $A_{1}(t_{c}^{-})$, $A_{2}(t_{c}^{-})$, 
and $\tilde u_{10}(x,t_{c})$. Additionally, we can  
take into account only the fast dependence of 
$\tilde u_{20}$ on $t$, i.e., the $t$ dependence that is contained 
in factors of the form $y=x-x_{20}-v_{d}t$.  
Denoting this approximation of $\tilde u_{20}(x,t)$ by $\bar u_{20}(y,t_{c})$, 
we obtain:
\begin{eqnarray} &&
\!\!\!\!\!\!\!\!\!\!\!
\Delta\phi_{1}(x,t_{c})\!=\!
-2\varepsilon_{2}A_{1}(t_{c}^{-})A_{2}(t_{c}^{-})\tilde u_{10}(x,t_{c}) \times
\nonumber \\&&
\!\!\!\int_{t_{c}-\Delta t_{c}}^{t_{c}+\Delta t_{c}} 
\!\!\!\!\!\!\!\! dt'  \bar u_{20}(x-x_{20}-v_{d}t',t_{c}). 
\label{rda5_add1}
\end{eqnarray}  
The integrand on the right hand side of Eq. (\ref{rda5_add1}) 
is sharply peaked at a small interval about $t_{c}$.  
Therefore, we can extend the limits of the integral to $-\infty$ and $\infty$. 
In addition, we change the integration variable from $t'$ to 
$y=x-x_{20}-v_{d}t'$ and obtain
\begin{eqnarray} &&
% 1-line version 1
\!\!\!\!\!\! 
\Delta\phi_{1}(x,t_c)=
-\frac{2\varepsilon_{2}A_{1}(t_{c}^{-})A_{2}(t_{c}^{-})}{|v_{d}|}\tilde u_{10}(x,t_{c}) 
\!\!\!\int_{-\infty}^{\infty} \!\!\!\!\!\!\!\! dy \, \bar u_{20}(y,t_{c}).
\nonumber \\&&
%\;\;\;\;\;\;\;\;\;\;\;\;\;\;
%\times
\label{rda6}
\end{eqnarray}    
The total collision-induced amplitude shift of pulse 1 $\Delta A_{1}^{(c)}$ 
is related to the collision-induced change in the concentration of the pulse  
$\Delta\phi_{1}(x,t_c)$ by:
\begin{eqnarray}&&
\!\!\!\!\!\!\!\!\!\!\!\!\!\!
\Delta A_{1}^{(c)}=\left[\int_{-\infty}^{\infty} \!\!\!\!\! dx \, 
\tilde u_{10}(x,t_{c})\right]^{-1}
\!\!\int_{-\infty}^{\infty} \!\!\!\!\! dx \, 
\Delta\phi_{1}(x,t_c)
\label{rda6_add1}
\end{eqnarray}       
(see Appendix \ref{appendA}). 
Substituting Eq. (\ref{rda6}) into  Eq. (\ref{rda6_add1}),  
we find that the total collision-induced amplitude shift of pulse 1 is: 
\begin{eqnarray} &&
\!\!\!\!\!\!\!\!
\Delta A_{1}^{(c)}=-\frac{2\varepsilon_{2}A_{1}(t_{c}^{-})A_{2}(t_{c}^{-})}{|v_{d}|}
\int_{-\infty}^{\infty} dy \, \bar u_{20}(y,t_{c}).
\label{rda7}
\end{eqnarray}    
Equation (\ref{rda7}) is expected to hold for generic pulse shapes $u_{j0}(x,t)$
with tails that exhibit exponential or faster than exponential decay.
Using Eq. (\ref{rda7}) for a fast collision between two Gaussian pulses 
with initial widths $W_{j0}$, we find that the collision-induced amplitude shift 
in this case is given by: 
\begin{eqnarray} &&
\!\!\!\!\!\!\!\!
\Delta A_{1}^{(c)}=-(8 \pi)^{1/2} \varepsilon_{2} W_{20}
A_{1}(t_{c}^{-}) A_{2}(t_{c}^{-})/|v_{d}|. 
\label{rda8}
\end{eqnarray}                   
In a similar manner, we find that the collision-induced 
amplitude shift in a fast collision between two square pulses with initial widths 
$W_{j0}$ is: 
\begin{eqnarray} &&
\!\!\!\!\!\!\!\!
\Delta A_{1}^{(c)}=-2 \varepsilon_{2} W_{20}
A_{1}(t_{c}^{-}) A_{2}(t_{c}^{-})/|v_{d}|. 
\label{rda8_add1}
\end{eqnarray} 
Equations (\ref{rda8}) and (\ref{rda8_add1}) are similar to 
Eqs. (\ref{coll8}) and (\ref{coll8_add1}) for the collision-induced 
amplitude shift in linear waveguides with weak cubic loss.     
Equations (\ref{rda8}) and (\ref{rda8_add1}) are also similar to 
Eq. (\ref{coll8_add2}) for the amplitude shift in a fast two-soliton collision 
in a nonlinear optical waveguide with weak cubic loss.  
Based on the similar forms of Eqs. (\ref{rda7})-(\ref{rda8_add1}) 
and Eq. (\ref{coll8_add2}) we conclude that pulses in linear physical systems 
described by the diffusion-advection model (\ref{rda1}) 
exhibit soliton-like behavior in fast collisions.

\subsection{Numerical simulations}
\label{diffusion_simu}
To check the predictions for soliton-like behavior in the collisions, 
we carry out numerical simulations with Eq. (\ref{rda1}). 
The equation is numerically solved by the split-step method 
with periodic boundary conditions \cite{Verwer2003}. 
For concreteness, we present here the results of simulations 
with parameter values $\varepsilon_{1}=0.01$  
and $\varepsilon_{2}=0.01$. The values of $v_{d}$ are varied 
in the intervals $-60 \le v_{d} \le -4$ and $4 \le v_{d} \le 60$.  
We carry out the simulations for two cases: (1) collisions between two Gaussian pulses;   
(2) collisions between two square pulses. The initial condition in the first case  
consists of two Gaussian pulses of the form (\ref{rda_IC1}) with parameter values  
$A_{j}(0)=1$, $W_{j0}=4$, $x_{10}=0$, and $x_{20}=\pm 20$. 
The initial condition in the second case  
consists of two square pulses of the form (\ref{rda_IC2}) with parameter values  
$A_{j}(0)=1$, $W_{j0}=4$, $x_{10}=0$, and $x_{20}=\pm 10$. 
The numerical and theoretical values of $\Delta A_{1}^{(c)}$ are calculated by the same 
methods that were used for the linear waveguide system.

\begin{figure}[htbp]
\centerline{\includegraphics[width=.95\columnwidth]{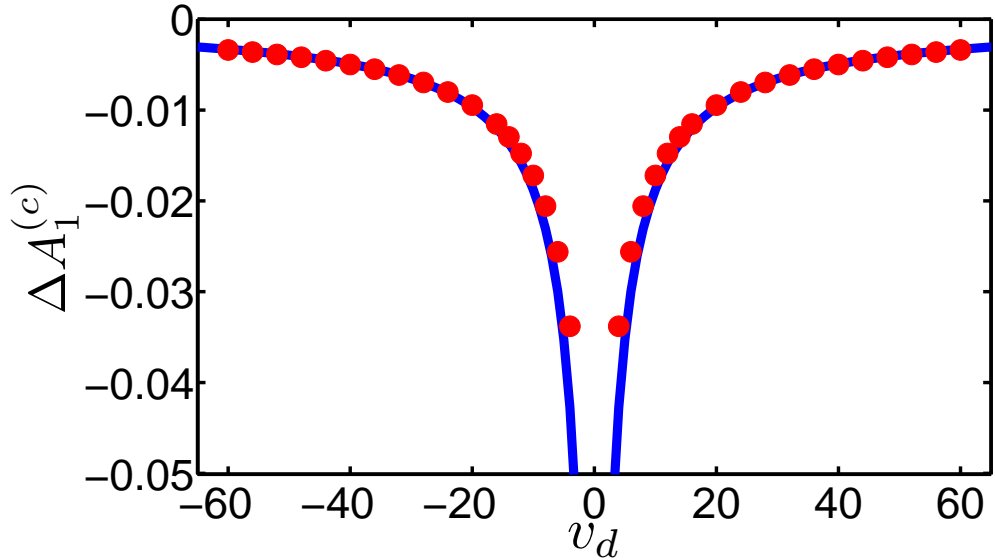}}
\caption{(Color online) The collision-induced amplitude shift of pulse 1 
$\Delta A_{1}^{(c)}$ vs advection velocity $v_{d}$ 
in a fast collision between two Gaussian pulses in a system, 
described by the diffusion-advection model (\ref{rda1}).  
The red circles correspond to the result obtained by simulations with Eq. (\ref{rda1}). 
The solid blue line corresponds to the prediction of Eq. (\ref{rda8}).}
\label{fig3}
\end{figure}

Figure \ref{fig3} shows the dependence of the 
collision-induced amplitude shift $\Delta A_{1}^{(c)}$ on $v_{d}$ 
for fast collisions between two Gaussian pulses, as obtained 
by simulations with Eq. (\ref{rda1}). 
The analytic prediction of Eq. (\ref{rda8}) is also shown. 
The agreement between the result of the simulations 
and the analytic prediction is very good. 
Indeed, the relative error is less than 9$\%$ for $|v_{d}|>10$ 
and less than 3$\%$ for $|v_{d}|>20$. 
Even at $|v_{d}| \simeq 4$, the relative error is only 20$\%$.

\begin{figure}[htbp]
\centerline{\includegraphics[width=.95\columnwidth]{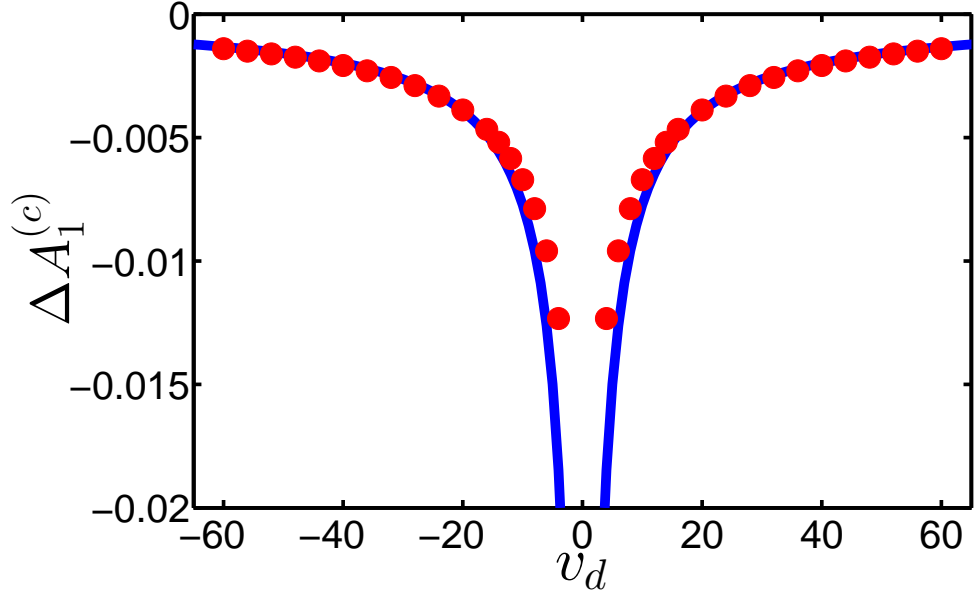}}
\caption{(Color online) 
The collision-induced amplitude shift of pulse 1 
$\Delta A_{1}^{(c)}$ vs advection velocity $v_{d}$ 
in a fast collision between two square pulses in a system, 
described by the diffusion-advection model (\ref{rda1}).  
The red circles correspond to the result obtained by simulations with Eq. (\ref{rda1}). 
The solid blue line corresponds to the prediction of Eq. (\ref{rda8_add1}).}
\label{fig4}
\end{figure}

The dependence of the collision-induced amplitude shift $\Delta A_{1}^{(c)}$ on $v_{d}$ 
for fast collisions between two square pulses is shown in Figure \ref{fig4}.  
Both the result obtained by simulations with Eq. (\ref{rda1}) 
and the analytic prediction of Eq. (\ref{rda8_add1}) are shown. 
The agreement between the result of the numerical simulations 
and the analytic prediction is very good. 
In particular, the relative error is less than 13$\%$ for $|v_{d}|>10$ 
and less than 6$\%$ for $|v_{d}|>20$. At $|v_{d}|\simeq 4$,  
the relative error is 33$\%$. 
Similar results to the ones presented 
in Figs. \ref{fig3} and \ref{fig4} are obtained for other 
physical parameter values and for other pulse shapes. 
Based on these results, we conclude that pulses in linear systems, described 
by diffusion-advection models, indeed exhibit soliton-like behavior in fast 
collisions in the presence of weak quadratic loss.

\section{Conclusions}
\label{conclusions}
We demonstrated that pulses of linear physical systems, 
weakly perturbed by nonlinear dissipation, 
exhibit soliton-like behavior in fast collisions. 
The behavior was demonstrated for linear waveguides 
with weak cubic loss and for systems described by 
linear diffusion-advection models with weak quadratic loss.
We showed that in both systems, the expressions for the collision-induced 
amplitude shifts due to the nonlinear loss have the same form as the expression 
for the amplitude shift in a fast collision between two optical solitons 
in a nonlinear optical waveguide with weak cubic loss. 
Our analytic predictions are confirmed by 
numerical simulations with the corresponding coupled   
linear evolution models with weak nonlinear loss.   
These results show that conclusions drawn from analysis of fast two-soliton 
collisions in the presence of weak dissipation can be applied for understanding 
the dynamics of fast two-pulse collisions in a large class 
of weakly perturbed linear physical systems, 
even though the pulses in the linear systems are not shape preserving. 
Furthermore, our results open the way for studying dynamics of fast collisions  
between pulses of weakly perturbed linear physical systems in an arbitrary spatial dimension, 
which is typically impossible for collisions between solitons in systems described by 
nonlinear Schr\"odinger models, due to the instability of the solitons in dimension 
higher than one.

\section*{Acknowledgments}
Q.M.N. and T.T.H are supported by the Vietnam National Foundation 
for Science and Technology Development (NAFOSTED) 
under Grant No. 101.99-2015.29.

\section*{Author contribution statement}
All authors contributed to this work equally.

\appendix
\section{Calculation of $\Delta A_{1}^{(c)}$ from $\Delta\Phi_{1}(t,z_{c})$ 
and $\Delta\phi_{1}(x,t_{c})$}
\label{appendA}
In this Appendix, we derive relations (\ref{coll6_add1}) and (\ref{rda6_add1}) 
between the collision-induced amplitude shift $\Delta A_{1}^{(c)}$ and the 
collision-induced changes in the envelopes of the electric field 
and in material concentration $\Delta\Phi_{1}(t,z_{c})$ and $\Delta\phi_{1}(x,t_{c})$. 
These relations were used to obtain Eq. (\ref{coll7}) and Eq. (\ref{rda7}) 
for $\Delta A_{1}^{(c)}$ from Eqs. (\ref{coll6}) and (\ref{rda6}), respectively.

We start by considering the linear waveguide system with weak linear and 
cubic loss, described by Eq. (\ref{coll1}). Employing the relation 
$\psi_{1}(t,z_{c})=\psi_{10}(t,z_{c})+\Delta\phi_{1}(t,z_{c})$, we obtain: 
\begin{eqnarray}&&
\!\!\!\!\!\!\!\!\!\!\!\!\!\!\!\!
\int_{-\infty}^{\infty} \!\!\!\!\!\!\!\! dt |\psi_{1}(t,z_{c})|^{2}=
\int_{-\infty}^{\infty} \!\!\!\!\!\!\!\! dt |\psi_{10}(t,z_{c})+\Delta\phi_{1}(t,z_{c})|^{2}.
\label{app1_collB1}
\end{eqnarray}  
Using the definitions of $\Psi_{10}$, $\tilde\Psi_{10}$, and $\Delta\Phi_{1}$, 
in Eq. (\ref{app1_collB1}), we obtain: 
\begin{eqnarray}&&
\!\!\!\!\!\!\!\!\!\!\!\!
\int_{-\infty}^{\infty} \!\!\!\!\!\!\!\! dt |\psi_{1}(t,z_{c})|^{2}\!=\!\!\!
\int_{-\infty}^{\infty} \!\!\!\!\!\!\!\! dt \left[A_{1}(z_{c}^{-})\tilde\Psi_{10}(t,z_{c})\!
+\!\Delta\Phi_{1}(t,z_{c})\right]^{2} 
\!\!\!\!. 
%\,
\label{app1_collB2}
\end{eqnarray}     
Expanding the integrand on the right hand side of Eq.  (\ref{app1_collB2}), 
while keeping only the first two leading terms, we arrive at: 
\begin{eqnarray}&&
\!\!\!\!\!\!\!\!\!
\int_{-\infty}^{\infty} \!\!\!\!\!\!\!\! dt |\psi_{1}(t,z_{c})|^{2}\simeq
C_{1}A_{1}^{2}(z_{c}^{-})
\nonumber \\&&
+2A_{1}(z_{c}^{-})
\int_{-\infty}^{\infty} \!\!\!\!\!\!\!\! dt \tilde\Psi_{10}(t,z_{c})\Delta\Phi_{1}(t,z_{c}),
\label{app1_collB3}
\end{eqnarray}      
where $C_{1}=\int_{-\infty}^{\infty} \!\!\! dt \tilde\Psi_{10}^{2}(t,z_{c})$ 
is a constant \cite{coll_conserved}. On the other hand, we can write:
\begin{eqnarray}&&
\!\!\!\!\!\!\!\!\!
\int_{-\infty}^{\infty} \!\!\!\!\!\!\!\! dt |\psi_{1}(t,z_{c})|^{2}\!=\!
\left(A_{1}(z_{c}^{-}) +\Delta A_{1}^{(c)}\right)^{2}
\int_{-\infty}^{\infty} \!\!\!\!\!\!\!\! dt \tilde\Psi_{10}^{2}(t,z_{c})
\nonumber \\&&
\simeq
C_{1}A_{1}^{2}(z_{c}^{-})+2C_{1}A_{1}(z_{c}^{-}) \Delta A_{1}^{(c)}.
\label{app1_collB4}
\end{eqnarray}     
Equating the right hand sides of Eqs. (\ref{app1_collB3}) and (\ref{app1_collB4}), we obtain: 
\begin{eqnarray}&&
\!\!\!\!\!\!\!\!\!
\Delta A_{1}^{(c)}=
\frac{1}{C_{1}}
\int_{-\infty}^{\infty} \!\!\!\!\!\!\!\! dt \tilde\Psi_{10}(t,z_{c})\Delta\Phi_{1}(t,z_{c}), 
\label{supp_collB5}
\end{eqnarray}     
which is the relation used to derive Eq. (\ref{coll7}) from Eq. (\ref{coll6}).

We now treat systems described by the 
coupled linear diffusion-advection model (\ref{rda1}). 
Using the relation $u_{1}(x,t_{c})=u_{10}(x,t_{c})+\Delta\phi_{1}(x,t_{c})$, 
we obtain:
\begin{eqnarray}&&
\!\!\!\!\!\!\!\!\!\!\!\!\!\!\!\!
\int_{-\infty}^{\infty} \!\!\!\!\!\!\!\! dx \, u_{1}(x,t_{c})=
\int_{-\infty}^{\infty} \!\!\!\!\!\!\!\! dx 
\left[u_{10}(x,t_{c})+\Delta\phi_{1}(x,t_{c})\right].
\label{app1_collB6}
\end{eqnarray}  
From the definition of $\tilde u_{10}(x,t_{c})$ it follows that 
\begin{eqnarray}&&
\!\!\!\!\!\!\!\!\!\!\!\!\!\!\!\!
\int_{-\infty}^{\infty} \!\!\!\!\!\!\!\! dx \, u_{1}(x,t_{c})=C_{2}A_{1}(t_{c}^{-})+
\int_{-\infty}^{\infty} \!\!\!\!\!\!\!\! dx \, \Delta\phi_{1}(x,t_{c}),
\label{app1_collB7}
\end{eqnarray}  
where $C_{2}=\int_{-\infty}^{\infty} \!\!\!\!\! dx \tilde u_{10}(x,t_{c})$ 
is a constant \cite{rda_conserved}. 
On the other hand, we can write:
\begin{eqnarray}&&
\!\!\!\!\!\!\!\!\!\!\!\!\!\!\!\!
\int_{-\infty}^{\infty} \!\!\!\!\!\!\!\! dx \, u_{1}(x,t_{c})=
\left(A_{1}(t_{c}^{-})+\Delta A_{1}^{(c)}\right)
\int_{-\infty}^{\infty} \!\!\!\!\!\!\!\! dx \, \tilde u_{10}(x,t_{c})
\nonumber \\&&
=C_{2}A_{1}(t_{c}^{-})+C_{2} \Delta A_{1}^{(c)}.
\label{app1_collB8}
\end{eqnarray}  
Equating the right hand sides of Eqs. (\ref{app1_collB7}) 
and (\ref{app1_collB8}), we obtain: 
\begin{eqnarray}&&
\!\!\!\!\!\!\!\!\!
\Delta A_{1}^{(c)}=
\frac{1}{C_{2}}
\int_{-\infty}^{\infty} \!\!\!\!\!\!\!\! dx \, \Delta\phi_{1}(x,t_{c}),
\label{supp_collB9}
\end{eqnarray}     
which is the relation used to derive Eq. (\ref{rda7}) from Eq. (\ref{rda6}).

\section{Procedures for calculating the values of $\Delta A_{1}^{(c)}$
from the analytic predictions and from numerical simulations}
\label{appendB}
Let us describe the procedures used for calculating the values of 
the collision-induced amplitude shift $\Delta A_{1}^{(c)}$
from the analytic predictions and from results of numerical simulations. 
For concreteness, we demonstrate the 
implementation of these procedures for a collision between 
two Gaussian pulses in linear optical waveguides with weak linear and cubic loss. 
The implementation for collisions between pulses with other shapes 
and for collisions in physical systems described by linear 
diffusion-advection models is similar.

The analytic prediction for $\Delta A_{1}^{(c)}$ is obtained by 
employing Eq. (\ref{coll8}). The values of $A_{j}(z_{c}^{-})$ are 
calculated by solving an approximate equation 
for the dynamics of $A_{j}(z)$ for a single pulse, 
propagating in the presence of first and second-order dispersion, 
linear loss, and cubic loss. More specifically, using an energy balance calculation 
for this single-pulse propagation problem, we obtain
\begin{eqnarray}&&
\!\!\!\!\!\!\!\!\!\!\!\!
\partial_{z}\int_{-\infty}^{\infty} \!\!\!\!\!\!dt|\psi_{j}|^{2}\!=\!
-2\epsilon_{1}\int_{-\infty}^{\infty} \!\!\!\!\!\!dt|\psi_{j}|^{2}
-2\epsilon_{3}\int_{-\infty}^{\infty} \!\!\!\!\!\!dt|\psi_{j}|^{4}.
\label{app2_coll1}
\end{eqnarray}       
We express the approximate solution of the propagation equation as 
$\psi_{j}(t,z)=A_{j}(z)\tilde\Psi_{j0}(t,z)\exp[i\chi_{j0}(t,z)]$,
where $\tilde\Psi_{j0}(t,z)\exp[i\chi_{j0}(t,z)]$ is the solution 
of the propagation equation in the absence of linear and cubic loss
with initial amplitude $A_{j}(0)=1$. Substituting the relation for 
$\psi_{j}(t,z)$ into Eq. (\ref{app2_coll1}), we obtain: 
\begin{eqnarray}&&
\!\!\!\!\!\!\!\!\!\!\!\!
\frac{d}{dz} 
\left[I_{2}A_{j}^{2}\right]= 
-2\epsilon_{1}I_{2}A_{j}^{2}-2\epsilon_{3}I_{4}(z)A_{j}^{4} \,,  
\label{app2_coll2}
\end{eqnarray}
where $I_{2}=\int_{-\infty}^{\infty} \!\!\!dt \tilde\Psi_{j0}^{2}(t,z)=
\int_{-\infty}^{\infty} \!\!\!dt \tilde\Psi_{j0}^{2}(t,0)$ and 
$I_{4}(z)=\int_{-\infty}^{\infty} \!\!\!dt \tilde\Psi_{j0}^{4}(t,z)$. 
For Gaussian pulses with initial width $W_{j0}$, we find $I_{2}=\pi^{1/2}W_{j0}$ 
and $I_{4}(z)=\pi^{1/2}W_{j0}^{3}/[2(W_{j0}^{4}+4z^{2})]^{1/2}$. 
Using these relations in Eq. (\ref{app2_coll2}), we arrive at
\begin{eqnarray}&&
\!\!\!\!\!\!\!\!\!\!\!\!
\frac{d}{dz} 
\left(A_{j}^{2}\right)+2\epsilon_{1}A_{j}^{2}= 
-\frac{2^{1/2}\epsilon_{3}W_{j0}^{2}A_{j}^{4}}
{(W_{j0}^{4}+4z^{2})^{1/2}} \,. 
\label{app2_coll3}
\end{eqnarray}
The solution of Eq. (\ref{app2_coll3}) on the interval $[0,z]$ is 
\begin{eqnarray}&&
\!\!\!\!\!\!\!\!\!\!\!\!
A_{j}(z)=\frac{A_{j}(0)e^{-\epsilon_{1}z}}
{\left[1+2^{1/2}\epsilon_{3}W_{j0}^{2}\tilde I(0,z)A_{j}^{2}(0)\right]^{1/2}} \,,  
\label{app2_coll4}
\end{eqnarray}  
where 
\begin{eqnarray}&&
\!\!\!\!\!\!\!\!\!\!\!\!
\tilde I(y_{1},y_{2})=\int_{y_{1}}^{y_{2}}
\frac{dy \, e^{-2\epsilon_{1}y}}
{(W_{j0}^{4}+4y^{2})^{1/2}} \,.
\label{app2_coll7}
\end{eqnarray}  
We use Eq. (\ref{app2_coll4}) for calculating the values of $A_{j}(z_{c}^{-})$. 
Substitution of these values into Eq. (\ref{coll8}) yields the analytic 
prediction for $\Delta A_{1}^{(c)}$.

To calculate $\Delta A_{1}^{(c)}$ from the simulations, 
we need to separate the collision-induced amplitude shift from the 
amplitude shift due to single-pulse propagation. 
The procedure that we adopt is a generalization of the method 
used in Refs. \cite{PNC2010} and \cite{PC2012} for calculating 
the collision-induced amplitude shift in two-soliton collisions in the presence of nonlinear loss. 
More specifically, we calculate the value of $\Delta A_{1}^{(c)}$ from the simulations 
by using $\Delta A_{1}^{(c)}=A_{1}(z_{c}^{+})-A_{1}(z_{c}^{-})$, 
where $A_{j}(z_{c}^{+})$ is the limit from the right of $A_{j}$ at $z_{c}$. 
The values of $A_{1}(z_{c}^{-})$ and $A_{1}(z_{c}^{+})$ are obtained 
by solving Eq. (\ref{app2_coll3}) on the intervals $[z_{1},z_{c}]$ and $[z_{c},z_{2}]$, 
where $z_{1}$ and $z_{2}$ are the distances at which the collision effectively 
starts and ends, respectively. We estimates these distances by 
$z_{1}=z_{c}-a/|d_{1}|$ and $z_{2}=z_{c}+a/|d_{1}|$, 
where $a>0$ is a constant of the same order of magnitude as $W_{j0}$. 
The expressions obtained in this manner are:  
\begin{eqnarray} &&
\!\!\!\!\!\!\!\!\!\!\!\!
A_{1}(z_{c}^{-})=\frac{A_{1}(z_{1})e^{-\epsilon_{1}z_{c}}}
{\left[e^{-2\epsilon_{1}z_{1}}
+2^{1/2}\epsilon_{3}W_{j0}^{2}\tilde I(z_{1},z_{c})
A_{1}^{2}(z_{1})\right]^{1/2}}, 
\label{app2_coll5}
\end{eqnarray} 
and 
\begin{eqnarray} &&
\!\!\!\!\!\!\!\!\!\!\!\!
A_{1}(z_{c}^{+})=\frac{A_{1}(z_{2})e^{-\epsilon_{1}z_{c}}}
{\left[e^{-2\epsilon_{1}z_{2}}
-2^{1/2}\epsilon_{3}W_{j0}^{2}\tilde I(z_{c},z_{2})
A_{1}^{2}(z_{2})\right]^{1/2}}. 
\label{app2_coll6}
\end{eqnarray}  
Thus, we obtain the values of $A_{1}(z_{c}^{-})$ and $A_{1}(z_{c}^{+})$ 
by using Eqs. (\ref{app2_coll5}) and (\ref{app2_coll6}) with values of 
$A_{1}(z_{1})$ and $A_{1}(z_{2})$, which are measured from the simulations.


\begin{thebibliography}{}

\bibitem{Zakharov84} S. Novikov, S.V. Manakov, L.P. Pitaevskii, 
and V.E. Zakharov, {\it Theory of Solitons: The Inverse Scattering Method}  
(Plenum, New York, 1984). 

\bibitem{Malomed89} Y.S. Kivshar and B.A. Malomed, 
Rev. Mod. Phys. {\bf 61}, 763 (1989). 

\bibitem{Agrawal2001} G.P. Agrawal, {\it Nonlinear Fiber Optics} 
(Academic, San Diego, CA, 2001).

\bibitem{Mollenauer2006} L.F. Mollenauer and J.P. Gordon,  
{\it Solitons in Optical Fibers: Fundamentals and Applications} 
(Academic, San Diego, CA, 2006).  

\bibitem{Horton96} W. Horton and Y.H. Ichikawa, 
{\it Chaos and Structure in Nonlinear Plasmas}  
(World Scientific, Singapore, 1996).  

\bibitem{MM98} L.F. Mollenauer and P.V. Mamyshev,
IEEE J. Quantum Electron. {\bf 34}, 2089 (1998). 

\bibitem{CP2005} Y. Chung and A. Peleg, Nonlinearity {\bf 18}, 1555 (2005).

\bibitem{PNC2010} A. Peleg, Q.M. Nguyen, and Y. Chung, 
Phys. Rev. A {\bf 82}, 053830 (2010).

\bibitem{Chi89} S. Chi and S. Wen, Opt. Lett. {\bf 14}, 1216 (1989).

\bibitem{Malomed91} B.A. Malomed, Phys. Rev. A {\bf 44}, 1412 (1991).

\bibitem{Kumar98} S. Kumar, Opt. Lett. {\bf 23}, 1450 (1998).  

\bibitem{P2004} A. Peleg, Opt. Lett. {\bf 29}, 1980 (2004).

\bibitem{NP2010} Q.M. Nguyen and A. Peleg,  J. Opt. Soc. Am. B {\bf 27}, 1985 (2010).  

\bibitem{PC2012} A. Peleg and Y. Chung, Phys. Rev. A {\bf 85}, 063828 (2012).

\bibitem{Tkach97} F. Forghieri, R.W. Tkach, and A.R. Chraplyvy,
 in {\it Optical Fiber Telecommunications III}, I.P. 
Kaminow and T.L. Koch, eds., (Academic, San Diego, CA, 1997), Chapter 8.

\bibitem{Agrawal89a} G.P. Agrawal, P.L. Baldeck, and R.R. Alfano, 
Phys. Rev. A {\bf 39}, 3406 (1989).

\bibitem{Agrawal89b} G.P. Agrawal, P.L. Baldeck, and R.R. Alfano, 
Opt. Lett. {\bf 14}, 137 (1989). 

\bibitem{PCG2003} A. Peleg, M. Chertkov, and I. Gabitov, 
Phys. Rev. E {\bf 68}, 026605 (2003).

\bibitem{PCG2004} A. Peleg, M. Chertkov, and I. Gabitov, 
J. Opt. Soc. Am. B {\bf 21}, 18 (2004).

\bibitem{SP2004} J. Soneson and A. Peleg, Physica D {\bf 195}, 123 (2004). 

\bibitem{Agrawal2007a} Q. Lin, O.J. Painter, and G.P. Agrawal, 
Opt. Express {\bf 15}, 16604 (2007).

\bibitem{Dimensions1} The dimensionless distance $z$ in Eq. (\ref{coll1})  is 
$z=Z/(2L_{D})$, where $Z$ is the dimensional distance, 
$L_{D}=\tau_{0}^{2}/|\tilde\beta_{2}|$ is the dispersion length,
and $\tau_{0}$ is the pulse width. 
The dimensionless time is $t=\tau/\tau_{0}$, where $\tau$ is time. 
$\psi_{j}=E_{j}/\sqrt{P_{0}}$, where $E_{j}$ is the 
electric field of the $j$th pulse and $P_{0}$ is peak power.   
$d_{1}=2(\tilde\beta_{12}-\tilde\beta_{11})\tau_{0}/|\tilde\beta_{2}|$, 
where $\tilde\beta_{1j}=1/v_{gj}$ and   
$v_{gj}$ is the group velocity of the $j$th pulse. 
$\epsilon_{1}=2\tau_{0}^{2}\tilde\rho_{1}/|\tilde\beta_{2}|$  
and $\epsilon_{3}=2P_{0}\tau_{0}^{2}\tilde\rho_{3}/|\tilde\beta_{2}|$, 
where $\tilde\rho_{1}$ and $\tilde\rho_{3}$ are  
the dimensional linear and cubic loss coefficients.




\bibitem{Dimensions2} The dimensionless coordinate $x$ in Eq. (\ref{rda1}) is 
 $x=X/x_{0}$, where $X$ is the dimensional coordinate, and $x_{0}$ is a reference pulse width.  
The dimensionless time is $t=\tau/\tau_{D}$, where $\tau$ is time, $\tau_{D}=x_{0}^{2}/D$, 
and $D$ is the diffusion coefficient.     
$u_{j}=U_{j}/\rho_{0}$, where $U_{j}$ is the 
concentration of substance $j$ and $\rho_{0}$ is the peak concentration.  
$v_{d}=x_{0}V_{d}/D$, where $V_{d}$ is the dimensional advection velocity.  
$\varepsilon_{1}=x_{0}^{2}\tilde\varepsilon_{1}/D$ and   
$\varepsilon_{2}=\rho_{0}x_{0}^{2}\tilde\varepsilon_{2}/D$, 
where $\tilde\varepsilon_{1}$ and $\tilde\varepsilon_{2}$ are the  
dimensional linear and quadratic loss coefficients. 


\bibitem{Verwer2003} W.H. Hundsdorfer and  J.G. Verwer, 
{\it Numerical Solution of Time Dependent Advection-Diffusion-Reaction Equations} 
(Springer, New York, 2003).      

\bibitem{coll_conserved} Since the integral $\int_{-\infty}^{\infty} \!\!\! dt \tilde\Psi_{10}^{2}(t,z)$ 
is conserved by the unperturbed linear propagation equation, we can write: 
$C_{1}=\int_{-\infty}^{\infty} \!\!\! dt \tilde\Psi_{10}^{2}(t,0)$.  

\bibitem{rda_conserved}
Since the integral $\int_{-\infty}^{\infty} \!\!\!\!\! dx \tilde u_{1}(x,t)$ 
is conserved by the unperturbed linear diffusion equation, we can write: 
$C_{2}=\int_{-\infty}^{\infty} \!\!\!\!\! dx \tilde u_{1}(x,0)$  

\end{thebibliography}
\end{document}